\documentclass[12pt]{article}

\newcommand{\beq}[1]{\begin{equation} \label{#1} }
\newcommand{\eeq}   {\end{equation}}
\newcommand{\Frac}[2]{\frac{\textstyle \mathstrut #1}
{\textstyle \mathstrut #2}}
\newcommand{\valpha}{\mbox{\boldmath $\alpha$}}
\newcommand{\vsigma}{\mbox{\boldmath $\sigma$}}

\begin{document}

\title{Spin-Orbit-Like Terms in Semileptonic Weak Hamiltonian}
\author{A.L.Barabanov\\
{\it Kurchatov Institute, 123182 Moscow, Russia}}
\date{}
\maketitle

\abstract{It is shown that new spin-orbit-like terms appear in the
effective nonrelativistic weak Hamiltonian for nucleon provided
that nuclear potential is taken into account. Arguments for their
considerable enhancement, in particular, in relativistic nuclear
model of Walecka are advanced.}
\bigskip

\bigskip

Effective weak Hamiltonian for nucleon is given by covariant
product of lepton current $J_{\lambda}$ and nucleon current
operator $\Gamma_{\lambda}$. To use nonrelativistic nucleon wave
functions one puts effective Hamiltonian to nonrelativistic form
by Foldy--Wouthuysen (FW) transformation (see, e.g.,
\cite{Bli73}). Nonrelativistic Hamiltonian is a power series in
$k_{\lambda}/M$, where $k_{\lambda}$ is the 4-vector of
transferred momentum, and $M$ is the nucleon mass. To describe
$\beta$-decay, where an energy release is about $10^{-3}M$, one
uses only zero order terms. Whereas first order terms are of
importance for muon capture, where an energy release is about
$10^{-1}M$, as well as for neutrino-induced reactions involving
transferred momenta of the same order of magnitude.

For the sake of definiteness we shall consider muon capture.
Before capture a muon is in the 1s state of mesic atom and is
described by the 4-component wave function
$\psi_{\mu}({\bf r}_{\mu},t)$, $E_{\mu}$ is the total energy of
muon. A final neutrino with momentum ${\bf k}_{\nu}$ and energy
$E_{\nu}=k_{\nu}$ is described by the 4-component wave function
$\psi_{\nu}( {\bf r}_{\nu},t)$.  Assuming that the weak
nucleon-lepton interaction is pointlike one obtains for the lepton
current
\beq{1}
J_{\lambda}=
i\psi^+_{\nu}({\bf r},t)\gamma_4\gamma_{\lambda}
(1+\gamma_5)\psi_{\mu}({\bf r},t).
\eeq
Then the effective relativistic Hamiltonian can be written in the
form
\beq{2}
H_W=\frac{G\cos\theta_C}{\sqrt{2}}
iJ_{\lambda}\Gamma_{\lambda}\tau_-,
\eeq
where $G$ is the weak-interaction coupling constant, $\theta_C$ is
the Cabibbo angle, and the lowering operator $\tau_-$ transforms a
proton into a neutron. The operator of the weak nucleon current is
given by (see, e.g., \cite{Muk77})
\beq{3}
\Gamma_{\lambda}=\gamma_4\left(
g_V\gamma_{\lambda}+
\frac{g_M}{2M}\sigma_{\lambda\rho}k_{\rho}-
g_A\gamma_{\lambda}\gamma_5-
i\frac{g_P}{m}k_{\lambda}\gamma_5\right).
\eeq
Here $\sigma_{\lambda\rho}=
(\gamma_{\lambda}\gamma_{\rho}-\gamma_{\rho}\gamma_{\lambda})/2i$,
$m$ is the muon mass, and
$k_{\lambda}=({\bf k}_{\nu},-i(E_{\mu}-E_{\nu}))$ is the
4-momentum transfer. The form factors of vector interaction $g_V$,
axial-vector interaction $g_A$, weak magnetism $g_M$, and induced
pseudoscalar interaction $g_P$ depend on
$k^2=k_{\lambda}k_{\lambda}$. We omit the contribution of the
second class currents (i.e. scalar and tensor terms).

One usually performs FW transformation for free nucleon describing
by Hamiltonian $H_N=M\beta+\valpha{\bf p}$. First order terms in
$1/M$ were first obtained in \cite{Fuj59}, whereas the second
order corrections $\sim 1/M^2$ were calculated in
\cite{Fri66,Oht66} (see also \cite{Con93,Bar99}). However, the
nucleons inside a nucleus are not, in fact, free.

Let us take the one-nucleon relativistic Hamiltonian in the form
\beq{4}
H=M\beta+\valpha{\bf p}+U(r)+H_W,
\eeq
where nucleon-nucleus potential $U(r)$ is assumed for simplicity
to be of a central type. According to FW procedure, $H-M\beta$ has
to be presented as the sum of even $\cal E$ and odd $\cal O$
parts. Then the nonrelativistic Hamiltonian takes the form
\beq{5}
H'=M+\mbox{$\cal E$}+\frac{\beta\mbox{$\cal O$}^2}{2M}-
\frac{[\mbox{$\cal O$},[\mbox{$\cal O$},\mbox{$\cal E$}]]}{8M^2}-
\frac{i[\mbox{$\cal O$},\dot{\mbox{$\cal O$}}]}{8M^2}+\ldots,
\eeq
where $[A,B]=AB-BA$.
It is well known that crossing of the operator $\valpha{\bf p}$
entering $\cal O$ with the potential $U(r)$, which belongs to
$\cal E$, in the term
$\sim [\mbox{$\cal O$},[\mbox{$\cal O$},\mbox{$\cal E$}]]$ leads
to Darwin and spin-orbit interactions being of the second order in
$1/M$. In a similar manner crossing of odd operators from $H_W$
with $U(r)$ gives additional second order terms, which never took
into account.

The result of FW transformation of the Hamiltonian (\ref{4}) is of
the form
\beq{6}
H'=M+\frac{{\bf p}^2}{2M}+U(r)+
\frac{\Delta U(r)}{8M^2}+
\frac{U'(r)}{4M^2r}(\vsigma[{\bf r}\times{\bf p}])+
H'_W,
\eeq
\beq{7}
\begin{array}{l}
H'_W=\Frac{G\cos\theta_C}{\sqrt{2}}\Bigl(iJ_4\Bigl[
G_V+G_P(\vsigma{\bf n}_{\nu})+{}
\\[\bigskipamount]
\phantom{H'_W=G\cos\theta_C}+g_A(\vsigma\Frac{{\bf p}}{M})+
g_P\Frac{iU'(r)}{4M^2r}(\vsigma{\bf r})\Bigr]+{}
\\[\bigskipamount]
\phantom{H'_W}+{\bf J}\Bigl[G_A\vsigma+g_V\Frac{{\bf p}}{M}+
g_V\Frac{U'(r)}{4M^2r}[\vsigma\times{\bf r}]\Bigr]\Bigr)
\tau_-,
\end{array}
\eeq
where ${\bf n}_{\nu}={\bf k}_{\nu}/k_{\nu}$. Nonrelativistic weak
Hamiltonian (\ref{7}) includes the known zero and first order
terms \cite{Muk77} and two new spin-orbit-like second order terms,
which are proportional to $U'(r)$. We omit all other second order
terms obtained in \cite{Fri66,Oht66}. Usual notations for
renormalized form factors are used
\beq{8}
\begin{array}{l}
G_V=g_V(1+\Frac{E_{\nu}}{2M}),\quad
G_P=\Frac{E_{\nu}}{2M}(g_P-g_A-g_V-g_M),
\\
G_A=g_A-\Frac{E_{\nu}}{2M}(g_V+g_M).
\end{array}
\eeq

It is worth noting now that nuclear spin-orbit coupling is
enhanced as compared to the term entering (\ref{6}) by the factor
of $\sim 20$ (and has the opposite sign) \cite{Boh69}. Thus, one
may hope for a similar enhancement of the spin-orbit-like terms in
the weak Hamiltonian (\ref{7}). It is of importance because the
second order terms enhanced by the factor of $\sim 20$ would be of
the same order of magnitude than  the first order terms.

To demonstrate the feasibility for such effect one can address to
relativistic nuclear model of Walecka \cite{Wal74,Ser86}. In its
simplest version a nucleon interacts with meson mean fields, one
of which, $\Phi(r)$, is scalar, and the other, $V(r)$, is timelike
component of 4-vector. The one-nucleon relativistic Hamiltonian
takes the form
\beq{9}
H=M\beta+\valpha{\bf p}+V(r)-\Phi(r)\beta+H_W.
\eeq
Both functions $V(r)$ and $\Phi(r)$ are positive, however, as it
is seen from (\ref{9}), $\Phi(r)$ and $V(r)$ represent attractive
and repulsive potentials, respectively. Both potentials are very
strong, e.g., $V(0)\simeq 0.37M$ and $\Phi(0)\simeq 0.45M$ for
$^{40}$Ca  nucleus \cite{Ser78}, but they almost cancel in
(\ref{9}).

This model describes the magnitude and the sign of nuclear
spin-orbit coupling \cite{Bro77}. Indeed, FW transformation of
(\ref{9}) gives
\beq{10}
\begin{array}{l}
H'=M+\Frac{{\bf p}^2}{2M}+V(r)-\Phi(r)+
\Frac{\Delta V(r)}{8M^2}-{}
\\[\bigskipamount]
{}-\Frac{\{\mbox{\boldmath $\nabla$},\{\mbox{\boldmath $\nabla$},
\Phi(r)\}\}}{8M^2}+
\Frac{V'(r)+\Phi'(r)}{4M^2r}(\vsigma[{\bf r}\times{\bf p}])+
H'_W,
\end{array}
\eeq
where $\{A,B\}=AB+BA$. It is seen that scalar and vector
contributions add up in the spin-orbit term, resulting in its
enhancement.

So, the question is, does the same summation arises for the
spin-orbit-like terms in $H'_W$? It turns out that this is the
case. The weak Hamiltonian is of the form (\ref{7}) with
$U'(r)\to V'(r)+\Phi'(r)$ if one neglects corrections $\sim\Phi/M$
to the form factors.

Thus, it is shown that new spin-orbit-like terms appear in the
effective nonrelativistic weak Hamiltonian for nucleon provided
that nuclear potential is taken into account. Being enhanced by
the factor of $\sim 20$, they may give the contribution of the
same order of magnitude than the first order terms, usually
allowed for. Finally, it is shown that such enhancement really
arises in the Walecka model.

New terms in weak Hamiltonian may result in an improvement of
description of some muon capture data. For instance, the results
obtained in \cite{Bru95} for $^{28}$Si nucleus point out an
anomalous low value of $g_P$.  On the other hand, search for new
terms contribution to the weak semileptonic interaction may be
considered as a test for relativistic nuclear model. The other
test based on a lowering of the threshold for $p\bar p$ production
on a nucleus was recently proposed in \cite{Hof99}.

\end{document}